\documentstyle[aps,twocolumn,prb]{revtex}
\input epsf
\begin{document}
\twocolumn[\hsize\textwidth\columnwidth\hsize\csname
@twocolumnfalse\endcsname

\draft

\title{Diffusion Quantum Monte Carlo Calculations of Excited States of
Silicon}

\author{A.~J.~Williamson\cite{address}, Randolph~Q.~Hood, R.~J.~Needs,
and G.~Rajagopal}

\address{Cavendish Laboratory, Madingley Road, Cambridge CB3 0HE, UK}

\date{\today}
\maketitle

\begin{abstract}
\begin{quote}
\parbox{16 cm}{\small The band structure of silicon is calculated at
the $\Gamma$, $X$, and $L$ wave vectors using diffusion quantum Monte
Carlo methods.  Excited states are formed by promoting an electron
from the valence band into the conduction band.  We obtain good
agreement with experiment for states around the gap region and
demonstrate that the method works equally well for direct and indirect
excitations, and that one can calculate many excited states at each
wave vector.  This work establishes the fixed-node DMC approach as an
accurate method for calculating the energies of low lying excitations
in solids.}

\end{quote}
\end{abstract}

]

\narrowtext 

\section{Introduction}
Electronic excitations play a crucial role in the physics and
chemistry of atoms, solids and molecules.  Calculating excitation
energies in large systems is a challenge for theoretical techniques
because an accurate description requires a realistic treatment of the
electron correlations.  The well-known Hartree-Fock (HF) method
includes exchange but not correlation effects, and therefore
overestimates band gaps and band widths by a large amount, while
Kohn-Sham density functional calculations within the local density
approximation (LDA) underestimate band gaps.  Here we report
a study of excitation energies in bulk silicon using the diffusion
quantum Monte Carlo (DMC) method~\cite{dmc,hammond}.  The DMC method
is very promising for applications to condensed matter because (i) it
explicitly includes electron-electron correlation effects, and (ii) it
scales reasonably well with system size, with the computational cost
increasing as the cube of the number of electrons.

It is now well established that DMC calculations can give an excellent
description of electron correlations in the ground state.  The range
of problems that could be addressed using DMC would be greatly
increased if one could also obtain accurate excitation energies.
Furthermore, the DMC method should be equally applicable to both
strongly and weakly correlated systems.  However, calculating
excitation energies in condensed matter systems is a formidable
challenge to DMC techniques because they are `$\frac{1}{N}$' effects,
i.e., the fractional change in energy is inversely proportional to the
number of electrons in the system.  The precision of the calculation
must therefore be sufficient to resolve this energy change amid the
statistical noise.  The system must also be large enough to give a
good description of the infinite solid, which is a severe constraint
for small band gap materials such as our test material, silicon.  So
far only a few DMC calculations of excitation energies in solids have
been reported.  In particular, we note the 8-atom simulation cell
calculations of an energy gap in a molecular nitrogen
solid~\cite{mitas1} and the ${\Gamma}_{25'} \! \rightarrow \! X_{1c}$
and ${\Gamma}_{1\nu} \! \rightarrow \!  {X}_{1c}$ excitations in
carbon diamond~\cite{mitas2}.  The excitations in these calculations
were indirect in reciprocal space, so that the excited state is
orthogonal to the ground state by translational symmetry.  In previous
work on solids~\cite{mitas1,mitas2,mitas3} it has been assumed that
the standard DMC method would give good results only for the lowest
energy state of each symmetry.  However, in this paper we will show
that the DMC method can be applied successfully to a wide range of
excitations in solids.

We have chosen silicon for our study because (i) DMC gives a good
account of the ground state~\cite{li}, (ii) electron correlations
significantly affect the band energies, and (iii) results from other
calculational methods, such as the LDA, HF~\cite{HF}, and
$GW$~\cite{Hedin,Hybertsen,Godby,Rohlfing} approximations are well
established, and a large amount of experimental data also exists.  In
this paper we explore the limits of the DMC method by calculating
excitations which are direct and indirect in reciprocal space, and
including several excitations at each wave vector.

It is important to distinguish between different types of excitation
energy.  In quasiparticle theory the quasiparticles correspond to the
poles of the one-particle Green function and are equal to the energies
for adding an electron to the system or subtracting one from it.  The
quasiparticle energies have both real and imaginary parts, the latter
giving the quasiparticle lifetime.  These energies are measured in
photoemission and inverse photoemission experiments. For the minimum
gap the imaginary part of the quasiparticle energy is zero and the
quasiparticle has an infinite lifetime.  In this case the
quasiparticle energy gap can be written as $E_g = E_{N+1} + E_{N-1} -
2 E_{N}$, where $E_{N+1}$, $E_{N-1}$, and $E_{N}$ are the ground state
total energies of the $N+1$, $N-1$ and $N$ electron systems.  This
energy gap is accessible within DMC methods, but we do not consider it
here.  In an optical absorption experiment a different process occurs
in which an electron is excited from the valence to the conduction
band.  In this case an exciton is formed and the lowest excitation
energy is smaller than $E_g$ by the exciton binding energy.  In the
calculations reported here we create excitonic states by exciting
electrons from a valence band state into a conduction band state.
Although the exciton binding energy is artificially increased by the
finite size of our simulation cell, it is still small (about 0.1 eV)
and therefore our results are comparable with theoretical
quasiparticle energies and experimental photoemission data.  We remark
that excitonic properties are of significant interest in their own
right and the ability to calculate both electron addition/subtraction
energies and excitonic energies is a significant advantage of the DMC
method.

This paper is organised as follows. In Sec. II we provide a brief
description of our calculational method. Section III is a detailed
discussion of the results of our study demonstrating the viability of
the DMC method for accurately determining the low lying excitation
energies in solids. We conclude with a summary in Section IV.

\section{Calculational Methods}
In the DMC method~\cite{dmc,hammond} imaginary time evolution of the
Schr\"{o}dinger equation is used to evolve an ensemble of
3$N$-dimensional electronic configurations towards the ground state.
Importance sampling is incorporated via a guiding wave function,
$\Phi$.  To make the calculations tractable we use the fixed node
approximation, in which the nodal surface of the wave function is
constrained to equal that of $\Phi$.  The fixed-node DMC method
generates the distribution $\Phi\Psi$, where $\Psi$ is the best
(lowest energy) wave function with the same nodes as $\Phi$. The
accuracy of the fixed node approximation can be tested on small
systems and normally leads to very satisfactory
results~\cite{hammond}.  We also use the short-time approximation for
the Green's function, whose effect can be tested and made very
small. We used a time step of 0.015 a.u., which gives small time-step
errors in silicon~\cite{li}.  The average number of configurations in
the ensemble was 384, and between 1220 and 2125 moves of all the
electrons in all the configurations were attempted, except for the
ground state where 3848 moves were attempted.  In all cases the
acceptance-rejection ratio was greater than 99.7\%.

We used a fcc simulation cell containing 16 silicon atoms, employing
periodic boundary conditions to reduce the finite size effects.  The
Si$^{4+}$ ions were represented by a norm-conserving non-local LDA
pseudopotential, and the non-local energy was evaluated using the
``locality approximation''~\cite{nl_dmc}.  The non-local potential was
sampled using the techniques of Fahy et al.~\cite{fahy_prb}  Our
guiding wave functions are of the Slater-Jastrow type:
\begin{equation}\label{hfjchi_trial}
\Phi = D^{\uparrow} D^{\downarrow} \exp \left[ \sum_{i=1}^{N}
\chi({\bf r}_i) - \sum_{i<j}^{N} u(r_{ij}) \right] \;\;\; ,
\end{equation}
where there are $N$ electrons in the simulation cell, $\chi$ is a
one-body function, $u$ is a two-body correlation factor which depends
on the relative spins of the two electrons, and $D^{\uparrow}$ and
$D^{\downarrow}$ are Slater determinants of up- and down-spin
single-particle orbitals.  We used a Fourier series expansion for the
$\chi$ function which was constrained to have the full symmetry of the
diamond structure and contains 6 free parameters.  The parallel- and
antiparallel-spin $u$ functions were constrained to obey the cusp
considitions~\cite{cusp} and contained a polynomial part with 11 free
parameters, as described in Ref.~\onlinecite{opt_prb}. The guiding
wave function contained a total of 28 parameters, whose optimal values
were obtained by minimizing the variance of the energy using 10$^5$
statistically independent electron
configurations~\cite{umrigar,opt_prb}, which were regenerated several
times during the minimization procedure.  The optimal parameter values
were obtained for the ground state wave function and were used for the
excited states as well.  Because the parameters occur only in the
nodeless Jastrow factor this procedure does not bias the DMC
excitation energies, which depend only on the nodal surfaces of the
guiding wave functions.

Our calculations are for many-body Bloch wave functions having
definite values of the crystal momentum or wave vector, $\bf k$.  The
Slater determinants were formed from the LDA orbitals calculated at
the $\Gamma$-point of the Brillouin zone of the simulation cell, which
unfolds to the $\Gamma$, $X$ and $L$ points of the primitive Brillouin
zone~\cite{note3}.  The wave vector of a determinant of these orbitals
is equal to one of the $\Gamma$, $X$ or $L$ wave vectors.  The
determinant for the ground state guiding wave function was constructed
from the valence band orbitals at the $\Gamma$, $X$ and $L$ points,
and has $\bf k$=0.  The excited state guiding wave functions were
formed by replacing an orbital in either the up- or down-spin
determinants of the ground state wave function by a conduction band
orbital.

The computational demands of DMC calculations are such that presently
it is not feasible to calculate excitation energies with cells larger
than our 16 atom cell, and therefore it is important to consider the
finite size effects.  The finite size effects can be divided into
``independent-particle finite size effects'' (IPFSE), which can be
modelled by LDA calculations, and ``Coulomb finite size effects''
(CFSE), which arise from the explicit use of Coulomb interactions
between the particles.  The HF, LDA and DMC ground state energies for
the 16 atom cell are $-103.67$, $-106.61$ and $-107.31(1)$ eV per
atom, respectively.  We estimate finite size errors in the ground
state calculations by performing calculations on simulation cells
containing 250 atoms (HF and LDA) and 128 atoms (DMC), giving
corrections of $-0.48$, $-1.31$ and $-1.02(3)$ eV per atom,
respectively. The IPFSE are therefore quite large, being about 21 eV
for the 16 atom cell used here.  However, LDA calculations performed
for a number of simulation cell sizes show that the cancellation
between the errors in the ground and excited states is so good that
the resulting finite size errors in the LDA excitation energies of the
16 atom cell are less than 0.1 eV.  We expect a similar cancellation
of errors in the DMC cancellations.  In our DMC simulations we have
also used a new formulation of the electron-electron interaction
within periodic boundary conditions~\cite{finite_size} which gives
much smaller CFSE than the standard Ewald form~\cite{ewald}.  However,
the CFSE also tend to cancel in the excitation energies and we
obtained almost identical results using our new interaction and the
Ewald interaction.

Each of the guiding wave functions we use contains some spin
contamination, i.e., they are not eigenstates of $\hat{S}^2$ but are
admixtures of different spin components. We have investigated the
effect of this spin contamination by calculating the $\Gamma_{25'} \!
\rightarrow \! X_{1c}$ excitation energy using a spin contaminated
single determinant and a two-determinant spin-singlet wave
function~\cite{note1}.  These calculations gave energies differing by
less than the statistical noise of 0.2 eV.  The exciton binding energy
in our calculations is enhanced because the exciton is confined to the
simulation cell. Following Ref.~\onlinecite{mitas1} we use the
Mott-Wannier formula to approximate the binding energy of the
localized exciton, giving 0.1 eV, which is smaller than the
statistical noise in our calculations.  Finally we investigated the
effect on the DMC energy of using single-particle orbitals which had
been relaxed by performing LDA calculations in the presence of the
excitation.  The resulting changes in the excitation energies were
smaller than the statistical noise.

\section{Results and Analysis}
In Table I we give our results for the 27 excitations studied,
together with HF, $GW$, and LDA data.  The characteristics of the HF,
$GW$, and LDA excitation energies are well known.  The $GW$
approximation gives extremely good excitation energies for weakly
correlated systems such as silicon, while the LDA excitation energies
are too small by 0.7-1.0 eV, and the HF excitation energies are much
too large.  The agreement between the DMC and $GW$ excitation energies
is good for the low energy excitations, but poorer for the higher
energy excitations.  The percentage, $\alpha_{ij}$, of the correlation
energy retrieved by our DMC calculation for the state formed by
exciting an electron from single-particle orbital $i$ to $j$ is
\begin{equation}\label{alpha_equation}
\alpha_{ij} = \frac{E_{DMC}^{ij}-E_{HF}^{ij}+\alpha_0
E_c^0}{E_{exact}^{ij}-E_{HF}^{ij}+E_c^0} \times 100 \;\;\;.
\end{equation}
\noindent The values of $E_{DMC}^{ij}$ and $E_{HF}^{ij}$ are given in
Table I.  We estimate the correlation energy for the ground state,
$E_c^0$, to be about -60 eV per simulation cell, and we estimate the
fraction of the ground state correlation energy retrieved by the DMC
calculation, $\alpha_0$, to be in the range 90-100\%.  For the
purposes of this comparison we use the $GW$ energies from
Ref.~\onlinecite{Rohlfing} for the $E_{exact}^{ij}$ because of the
incompleteness and uncertainty of the experimental data.  We note that
the various $GW$ calculations for
silicon~\cite{Hybertsen,Godby,Rohlfing} are in excellent agreement
with one another and are also in good agreement with the available
experimental data.  The experimental data are reported as band
energies, so that to form the excitation energies we have to take
differences between the experimental values, which adds to the
uncertainties.  When we present our results for band energies we will
compare with the experimental data, which is given in Table II.

The values of $\alpha_{ij}$ given by this analysis slowly decrease
with increasing excitation energy, so that for the largest excitation
energies the $\alpha_{ij}$ are 2-3 percentage points smaller than for
the smallest excitation energies.  This conclusion is insensitive to
the values of $E_c^0$ and $\alpha_0$.  This suggests the following
rationale for our results.  The HF excitation energies are much too
large because the correlation energy is neglected.  The fraction of
the correlation energy retrieved by our DMC calculations slowly
decreases with increasing excitation energy.  However, because the
contribution of the correlation increases rapidly with increasing
excitation energy the DMC excitation energies are somewhat too large.
This analysis indicates that the residual errors in the DMC excitation
energies are mostly due to the errors in the nodal surfaces of the
excited state guiding wave functions rather than in the ground state,
and that the quality of the nodal surfaces falls with increasing
excitation energy.

To study whether the DMC method works for direct excitations we
considered pairs of excitations where a single-particle orbital with a
particular wave vector is removed from the determinant and replaced
(i) by a higher energy orbital at the same wave vector and (ii) by an
equivalent higher energy orbital at a different wave vector.  For
example, when calculating the $X_{4} \! \rightarrow \! X_{1c}$
excitation energy, we replace an $X_{4}$ orbital at a particular
$X$-point by an $X_{1c}$ orbital at the same $X$-point to give a
direct excitation, while for the $X_{4} \! \rightarrow \! X_{1c}^*$
excitation we replace the $X_{4}$ orbital with an $X_{1c}$ orbital
from a different $X$-point, giving an indirect excitation.  We have
investigated three such direct-indirect excitation pairs and found
reasonable consistency between the results.  We have calculated a
total of 6 direct ($\Gamma$-point) excitations and the level of
accuracy is indistinguishable from that for indirect excitations,
showing that the DMC method can be applied to direct excitations.  In
addition we have successfully calculated a total of 10 excitation
energies at the $X$-point and 11 at the $L$-point, which demonstrates
that the DMC method works for higher excitations as well.

The 27 entries in Table I correspond to transitions between 7 valence
and 5 conduction band energy levels.  To obtain DMC band energies,
$\epsilon_i$, we performed a least squares fit to the DMC data by
minimizing $\Sigma [E_{DMC}^{ij} - (\epsilon_i - \epsilon_j)]^2$ with
respect to the $\epsilon_i$, where the sum is over the 27 excitation
energies, $E_{DMC}^{ij}$, listed in Table I.  The resulting band
energies are given in Table II together with other theoretical data
and with experimental data.  For greater clarity the DMC band energies
are plotted in Fig. 1 together with an empirical pseudopotential band
structure~\cite{chelikowsky} which is in good agreement with the
available experimental data.  The energies at the top of the valence
band have been aligned.  The DMC band energies are very much better
than the HF values because of the inclusion of correlation effects.
For the lower part of the valence band the DMC energies lie
consistently 1 to 1.5 eV below the empirical pseudopotential, $GW$,
and experimental data.  The DMC band energies around the gap region
are in good agreement with the empirical pseudopotential, $GW$, and
experimental data.

The success of our excitation energy calculations is very encouraging.
To appreciate why our calculations are so successful we must consider
the fixed-node DMC algorithm in more detail.  First we consider ground
state calculations.  The fixed-node DMC energy is a variational bound
on the exact ground state energy, and if the nodal surface of the
guiding wave function is exact then the resulting fixed-node DMC
energy is also exact~\cite{note2}. The exact ground state wave
function ``tiles the configuration space'', which means that all nodal
pockets are related by permutation symmetry~\cite{nodes}.  Normally
one chooses guiding wave functions which also have the tiling property
and consequently the DMC simulation may be performed in any subset of
the nodal pockets.

For excited states the situation is somewhat different.  If the
fixed-node constraint could be removed the algorithm would, in the
limit of a long simulation, give the ground state energy.  However,
the key point is that the fixed-node approximation prevents this
collapse and allows us to obtain good approximations for excited state
energies.  One can readily show that if the nodal surface of the
guiding wave function is exactly that of the $n$th eigenstate, then
the fixed-node DMC procedure gives the exact energy of the $n$th
eigenstate, regardless of whether the guiding wave function overlaps
with lower energy states. However, if we use a guiding wave function
that does not have the exact nodal surface of the excited state we are
modeling, the resulting DMC energy may not be above the exact energy
of that state because of the possibility of mixing in lower energy
states with the same symmetry.

Another complication arises from the fact that the nodal pockets of
approximate excited state wave functions may be inequivalent.
Therefore, as the DMC simulation proceeds, the population of
configurations in some nodal pockets will dominate over others, and in
principle the results could depend on which of the pockets were
occupied at the start of the simulation.  An illustration of this type
of behaviour for the first excited state of an electron in an infinite
square well is given on page 186 of Ref.~\onlinecite{hammond}, which
shows the DMC energy decreasing linearly with the magnitude of the
error in the nodal position.

Although there are significant differences between ground and excited
state fixed-node DMC calculations, the criterion for obtaining good
energies is the same, i.e., the nodal surface of the guiding wave
function must be of good quality.  Our results demonstrate that the
nodal surfaces from determinants of LDA orbitals are fully adequate
for calculating excitation energies around the gap region in silicon,
but give poorer results for higher excitation energies.  It will
therefore be necessary to optimize the nodal surfaces of the guiding
wave functions to obtain more accurate estimates of the higher
excitation energies.  We note that a DMC calculation by Grimes {\em et
al.}~\cite{grimes} for an excited state of the hydrogen molecule with
the same symmetry as the ground state also gave a good excitation
energy.  This result lends further support to our contention that the
fixed-node DMC method can be applied to a wide range of excited states.

\section{Summary}
In conclusion, the DMC method is a stable and accurate method for
calculating low lying excitation energies in solids.  The fixed-node
approximation works to our advantage by preventing collapse to lower
energy states.  The accuracy of the excited state energies is
determined by the quality of the nodal surfaces of the guiding wave
functions.  We have obtained good values for the low lying excitation
energies, including direct and indirect excitations, and including
several excitations at each wave vector, which indicates that the
nodal surfaces of our guiding wave functions are of good quality.  We
have demonstrated that DMC calculations can be used to calculate
direct excitations, which is important because it allows us to obtain
excitation energies when there is no underlying translational
symmetry.  The fixed-node DMC method provides a unified framework for
calculating accurate ground and excited state energies.

\begin{center}
\bf{ACKNOWLEDGEMENTS}
\end{center}

We thank Matthew Foulkes for helpful conversations.  Financial support
was provided by the Engineering and Physical Sciences Research Council
(UK).  Our calculations are performed on the CRAY-T3D at the Edinburgh
Parallel Computing Centre, and the Hitachi SR2201 located at the
Cambridge HPCF. Guna Rajagopal acknowledges support from Hitachi Ltd.

\begin{table}[h]
\begin{tabular}{llrrrr}
Excitation & $\bf k$ & DMC & HF\protect\cite{HF} & $GW$
\protect\cite{Rohlfing} & LDA \\ \hline

$\Gamma_{1v} \! \rightarrow \! \Gamma_{2'}$ & $\Gamma$ & 18.05 & 27.9 & 15.84 & 15.14 \\
$\Gamma_{1v}\! \rightarrow \! \Gamma_{15}$ & $\Gamma$ & 17.38 & 26.9 & 15.31 & 14.50 \\
$\Gamma_{1v} \! \rightarrow \! L_{1c}$ & $L$ & 16.27 & 25.4 & 14.14 & 13.39 \\
$\Gamma_{1v} \! \rightarrow \! X_{1c}$ & $X$ & 14.93 & 24.2 & 13.38 & 12.58 \\
$\Gamma_{25'} \! \rightarrow \! L_{3}$ & $L$ & 4.75 & 8.7 & 4.25 & 3.31 \\
$\Gamma_{25'} \! \rightarrow \! \Gamma_{15}$ & $\Gamma$ & 3.82 & 8.0 & 3.36 & 2.55 \\
$\Gamma_{25'} \! \rightarrow \! L_{1c}$ & $L$ & 2.35 & 6.5 & 2.19 & 1.44 \\
$\Gamma_{25'} \! \rightarrow \! X_{1c}$ & $X$ & 1.34 & 5.3 & 1.43 & 0.63 \\
$X_{1v} \! \rightarrow \! \Gamma_{15}$ & $X$ & 12.42 & 20.5 & 11.31 & 10.36 \\
$X_{1v} \! \rightarrow \! X_{1c}^*$    & $L$ & 10.37 & 17.8 & 9.38 & 8.44 \\
$X_4 \! \rightarrow \! \Gamma_{15}$    & $X$ & 6.91 & 12.7 & 6.29 & 5.39 \\
$X_4 \! \rightarrow \! \Gamma_{2'}$    & $X$ & 7.92 & 13.7 & 6.82 & 6.03 \\
$X_4 \! \rightarrow \! L_{1c}$         & $L$ & 5.71 & 11.2 & 5.12 & 4.28 \\
$X_4 \! \rightarrow \! X_{1c}$         & $\Gamma$ & 5.12 & 10.0 & 4.36 & 3.47 \\
$X_4 \! \rightarrow \! X_{1c}^*$       & $L$ & 4.86 & 10.0 & 4.36 & 3.47 \\
$L_{2'} \! \rightarrow \! L_{3}^*$     & $X$ & 15.60 & 24.1 & 13.95 & 12.92 \\
$L_{2'} \! \rightarrow \! \Gamma_{15}$ & $L$ & 14.75 & 23.4 & 13.06 & 12.16 \\
$L_{1v} \! \rightarrow \! L_{3}^*$     & $X$ & 12.29 & 19.8 & 11.39 & 10.30 \\
$L_{1v} \! \rightarrow \! \Gamma_{15}$ & $L$ & 11.55 & 19.1 & 10.50 & 9.54 \\
$L_{1v} \! \rightarrow \! L_{1c}$      & $\Gamma$ & 10.80 & 17.6 & 9.33 & 8.43 \\
$L_{1v} \! \rightarrow \! L_{1c}^*$    & $X$ & 9.87 & 17.6 & 9.33 & 8.43 \\
$L_{3'} \! \rightarrow \! \Gamma_{2'}$ & $L$ & 6.00 & 11.0 & 5.14 & 4.38 \\
$L_{3'} \! \rightarrow \! L_{3}^*$     & $X$ & 5.76 & 10.7 & 5.50 & 4.50 \\
$L_{3'} \! \rightarrow \! \Gamma_{15}$ & $L$ & 4.96 & 10.0 & 4.61 & 3.74 \\
$L_{3'} \! \rightarrow \! L_{1c}$      & $\Gamma$ & 3.90 & 8.5 & 3.44 & 2.63 \\
$L_{3'} \! \rightarrow \! L_{1c}^*$    & $X$ & 3.82 & 8.5 & 3.44 & 2.63 \\
$L_{3'} \! \rightarrow \! X_{1c}$      & $L$ & 2.83 & 7.3 & 2.68 & 1.82 \\

\end{tabular}

\caption{Excitation energies in eV calculated with the DMC, HF, $GW$,
and LDA methods.  The wave vector of the excited state wave function
is denoted by $\bf k$.  The statistical error bars on the DMC energies are
$\pm$0.2 eV.}

\label{promotion_results}
\end{table}

\begin{table}[h]
\begin{tabular}{lrrrrrc}

Band & DMC$^a$ & HF$^b$ & GW$^c$ & LDA$^a$ & Emp$^d$ & Experiment$^e$ \\ \hline
$\Gamma_{25'}$ & 0.00   & 0.0   & 0.00   & 0.00   & 0.00   & 0.00 \\
$\Gamma_{15}$  & 3.70   & 8.0   & 3.36   & 2.55   & 3.42   & 3.40,3.05 \\
$\Gamma_{2'}$  & 4.57   & 9.0   & 3.89   & 3.19   & 4.10   & 4.23,4.1 \\
$\Gamma_{1}$   & -13.58 & -18.9 & -11.95 & -11.95 & -12.36 & -12.5$\pm$0.6 \\
$X_{1c}$       & 1.51   & 5.3   & 1.43   & 0.63   & 1.17   & 1.25 \\ 
$X_{4}$        & -3.35  & -4.7  & -2.93  & -2.84  & -2.86  & -3.3$\pm$0.2,-2.9 \\
$X_{1v}$       & -8.79  & -12.5 & -7.95  & -7.81  & -7.69  & \\
$L_{1c}$       & 2.51   & 6.5   & 2.19   & 1.44   & 2.23   & 2.4$\pm$0.15,2.1 \\
$L_{3}$        & 4.55   & 8.7   & 4.25   & 3.31   & 4.34   & 4.15$\pm$0.1 \\ 
$L_{3'}$       & -1.32  & -2.0  & -1.25  & -1.19  & -1.23  & -1.2$\pm$0.2,-1.5 \\ 
$L_{1v}$       & -7.81  & -11.1 & -7.14  & -6.99  & -6.96  & -6.7$\pm$0.2 \\
$L_{2'}$       & -11.05 & -15.4 & -9.70  & -9.61  & -9.55  & -9.3$\pm$0.4 \\

\end{tabular}
\caption{Band energies of silicon in eV. a- This work. b-
Ref. \protect\cite{HF}.  c- Ref. \protect\cite{Rohlfing}. d-
Ref. \protect\cite{chelikowsky}. e- From the compilation given in
Ref. \protect\cite{Rohlfing}.  The statistical error bars on the DMC
energies are $\pm$0.2 eV.}

\end{table}

\begin{figure}
\begin{center}
\epsfysize=8cm \epsfbox{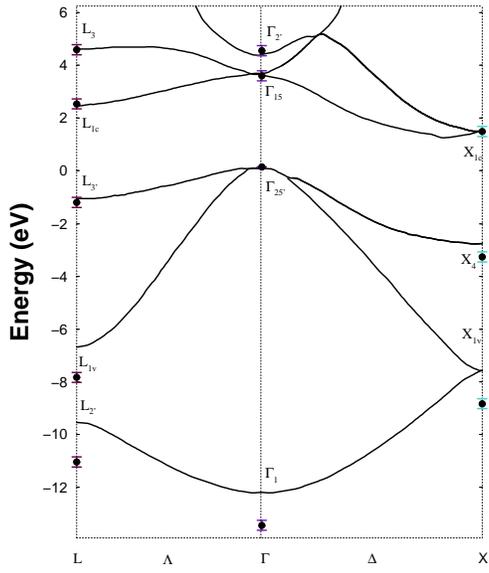}
\end{center}
\caption{The DMC band structure (filled circles with error bars).  As a guide
to the eye we also show empirical pseudopotential
data\protect\cite{chelikowsky} (solid lines).}
\label{bandstructure}
\end{figure}


\begin{thebibliography}{10}
\bibitem{note3} The LDA calculations were performed using a plane wave
basis set including all waves up to a cutoff energy of 15 Ry.
\bibitem{note2} We neglect errors from the short time approximation.
\end{thebibliography}
\end{document}